\documentclass[showpacs,twocolumn,prl,superscriptaddress]{revtex4}
\usepackage{amsmath}
\usepackage{graphicx}
\usepackage{dcolumn}
\usepackage{color}
\usepackage{bm}

\begin{document}
\title{Decoherence-assisted transport and quantum criticalities}

\author{Gian Luca Giorgi} 
\affiliation{Department of Physics, University College Cork, Cork,
  Republic of Ireland}

\author{Thomas Busch} \affiliation{Department of Physics, University
  College Cork, Cork, Republic of Ireland}
\affiliation{Quantum Systems Unit, Okinawa Institute of Science and
  Technology, Okinawa 904-0411, Japan}


\begin{abstract}
  We study the dynamics of a two-level quantum system interacting with
  an external environment that takes the form of an XY spin chain in
  the presence of an external magnetic field. While the presence of
  the bath itself can enhance the transition probability from the
  lower level to the upper level of the system, we show that this
  noise-assisted phenomenon is sensitive to a change of the quantum
  phase of the environment.  The derivative of the transition
  probability displays a maximum in correspondence with the critical
  value of the applied field both in the case of isotropic and
  anisotropic chains.
\end{abstract}

\maketitle

The unavoidable interaction of any realistic quantum system with its
environment is responsible for decoherence and dissipation, which are
usually considered detrimental effects for the implementation of
quantum information and quantum communication
tasks~\cite{nielsen,breuer}. However, recent experimental studies have
shown that some photosynthetic complexes exhibit long excitonic
coherence despite the interaction with their own
environment~\cite{engel2007,panit2010,collini2010}. Furthermore, the
possibility of decoherence-assisted quantum transport was recently
discovered in a series of theoretical Hamiltonian models.  Therefore,
fully understanding the role played by the noise in assisting quantum
transport phenomena has assumed a crucial importance and has become a
rich subject of investigation (see, for instance,
Refs.~\cite{plenio08,olaya08,mohseni08,rebentrost09,caruso09,
  sarovar,semiao,campos11,kassal12}).

While a big part of the theoretical investigations in this field deals
with quantum systems in contact with bosonic environments, coherent
transport induced by structured spin baths has for the first time been
taken into consideration in Ref.~\cite{sinayskiy}. There, it was shown
that, the transition probability from the lower to the upper level of a
two-level system can be increased even for finite temperatures when
modeling the environment as a series of spins $1/2$ arranged in a star
configuration. The authors also proved that the use of two separated
baths, correlated with each other, can help to improve the transport
compared to the use of a single environment.

At the same time, it is known that decoherence is sensitive to quantum
critical changes in the bath and that the correlation length of the
environment can influence the rate of loss of quantumness in a
system~\cite{paganelli,quan,rossini07,rossini2008,ions,damski}. A
possible justification for this behavior can be found in the
monogamous nature of entanglement \cite{dawson}.

Starting from the results of Ref.~\cite{sinayskiy}, here we take a
step forward by considering the case of a spin bath that can
experience a quantum phase transition and studying how a critical
change in the environment will influence the noise-induced properties
of a two-level system (TLS) coupled to the bath itself. The goal of
our work is to find a direct link between the phase transition of the
bath and the internal transition probability of the TLS.

We approach the problem by first considering the case of a TLS coupled
to a homogeneous XX chain in the presence of a transverse field. If
the system is coupled to the total magnetization of the bath, the
problem is exactly solvable. We show that the transition probability
inside the TLS is maximally sensitive to a change in the value of the
applied field near its critical point and justify the result by noting
that the bath renormalizes the system Hamiltonian parameters. The same
qualitative behavior can be observed in the case of an inhomogeneous
XY chain, where the analytical solution is no longer available, and
where the simple mechanism of renormalization cannot be applied any
more.

Let us consider a  TLS described by the Hamiltonian 
\begin{equation}
H_{S}=\frac{\Delta}{2} \sigma_{z}+T\sigma_{x}.
\end{equation}
Because of the detuning, $\Delta$, between the states
$|\uparrow\rangle$ and $|\downarrow\rangle$, a state initially
prepared in $|\downarrow\rangle$ has a maximum probability of
transition to $|\uparrow\rangle$ of
$P_{tr}=T^{2}/(T^{2}+\Delta^{2}/4)\le 1$, with perfect transition for
$\Delta=0$. As discussed by Sinayskiy \textit{et al.} in
Ref.~\cite{sinayskiy}, under certain circumstances, the presence of a
single spin bath or two different, correlated baths, can help increase
the maximum transition probability for finite values of $\Delta$.

Here our aim is to analyze the effect an internally structured
environment can have on a decoherence-assisted transition process. We
will consider
a spin environment and study its influence as a function of an
internal parameter which drives the system through a quantum critical
point. For this, we model the bath as an XY chain in the presence of a
transverse field, which can be described as
\begin{equation}
  H_{XY}=-J\sum_{l=1}^N\left[(\frac{1+\gamma}{2})\sigma_l^x\sigma_{l+1}^x
                         +(\frac{1-\gamma}{2})\sigma_l^y\sigma_{l+1}^y\right]
         -h M_z,
\label{hxy}
\end{equation} 
where the total magnetization along $z$ is $M_z=\sum_{l=1}^{N}
\sigma_{l}^{z}$ and where the boundary conditions are periodic
$\sigma_{N+1}^{\alpha}=\sigma_{1}^{\alpha}$. From now on we will take
$J=1$ and use it as a scale of energy.
In the thermodynamic limit the Hamiltonian $H_{XY}$ possesses a
critical point which can be observed when considering the average
magnetization per spin, $m_z=M_z/N$~\cite{barouch}
\begin{equation}
  \lim_{N \to \infty}m_z=\frac{1}{\pi}\int_0^\pi
                     \frac{\tanh \left[\beta\Lambda(h)/2\right]}{\Lambda(h)}
                                       (h-\cos\phi)\;d\phi,
  \label{magn}
\end{equation} 
where $\beta$ is the inverse temperature and $\Lambda(h)=\left[
  (h-\cos\phi)^2+\gamma^2 \sin^2 \phi \right]^{1/2}$. For finite
temperatures, the derivative of $m_z$ with respect to $h$ shows a
maximum for $h<h_c$, which turns into a divergence at $h_c=1$ for
$\beta \to \infty$.

In the following, we will first pay special attention to the case of
an isotropic chain, characterized by $\gamma=0$, and indicate the
Hamiltonian operator as $H_{XX}$.  For the sake of simplicity and
without any loss of generality, we will only consider non-negative
fields ($h \ge 0$) and note that $M_z$ is a conserved quantity since
$\left[H_{XX}, \sum_{l} \sigma_{l}^{z}\right]=0$.  For $h>h_c=1$, the
ground state is completely ordered along the direction $z$ and no
spin-spin correlations are present. Lowering $h$ below the critical
value $h_c$, the ground state starts acquiring magnons (flipped spins)
and for $h=0$ the number of magnons is exactly equal to $N/2$.  In a
pictorial representation, exploiting the Jordan-Wigner mapping between
spins and fermions, the transverse field $h$ plays the role of a
chemical potential and determines the value of the Fermi level: all
the levels with energy less than $h$ will belong to the Fermi sea and
will be filled, while all the levels with energy greater than $h$ will
be left empty. This mechanism implies that, for finite-size systems,
moving the field from zero to one, $N/2$ transition points between
different symmetry sectors of the Hamiltonian are crossed. In the
thermodynamic limit, this leads to an infinite-order quantum phase
transition, named Berezinsky-Kosterlitz-Thouless, which takes place
without spontaneous symmetry breaking.
 
We assume that the interaction between the bath and the system is
given by a coupling between the upper level $|\uparrow\rangle$ and the
total magnetization of the chain~\cite{sinayskiy}
 \begin{equation}
   H_{I}=-\Gamma |\uparrow\rangle\langle\uparrow | \otimes m_z.
\end{equation} 
Since $m_z$ is a constant of motion, one can see that
$[H_{XX},H_{I}]=0$ and therefore the degrees of freedom of the
environment can be eliminated. As a consequence, an exact solution for
the dynamics of the system can be obtained. Notice that we are not
dealing with a purely dephasing interaction, which would be
characterized by $[H_S,H_{I}]=0$, but that our model is suitable to
describe a fully dissipative dynamical evolution.

Assuming that at $t=0$ the system and the bath are in a product state
[$\rho(0)=\rho_S (0)\otimes\rho_B(0)$], the density matrix of the
system will evolve according to
\begin{equation}
\rho_{S} (t)={\rm Tr}_B [ e^{-i (H_S+H_{I})t}\rho_S (0)\otimes \rho_B(0) e^{i (H_S+H_{I}) t} ],
\end{equation}
where we have now eliminated $H_{XX}$ by exploiting $[H_{XX},H_{I}]=0$
and the invariance of the trace under cyclic permutations. Even if
formally eliminated from the dynamics, $H_{XX}$ is present through the
initial state of the bath $ \rho_B(0)$, which will be assumed to be in
a thermal distribution $ \rho_B(0)=e^{-\beta H_{XX}}/Z$ at the inverse
temperature $\beta$, where the partition function is $Z={\rm Tr
}[e^{-\beta H_{XX}}] $.

Both $H_{XX}$ and $H_{I}$ can be written in a diagonal form by using
the aforementioned Jordan-Wigner transformation, which maps spins into
spinless fermions~\cite{lieb}.
Labelling the eigenstates of the bath $|\lambda\rangle$ and the
corresponding eingevalues $\lambda$, the matrix elements $\langle i
|\rho_S| j\rangle=\rho_S^{ij}$ will evolve in time according to
\begin{equation}
  \rho_{S} ^{ij}(t)=\frac{1}{Z}\sum_{\lambda}   e^{-\beta \lambda}    \langle i\vert e^{-i H_S^{(\lambda)}t}\rho_S (0) e^{i H_S^{(\lambda)} t} \vert j \rangle,\label{wei}
\end{equation}
with $H_{S}^{(\lambda)}=H_S-\Gamma |\uparrow\rangle\langle\uparrow |
\langle\lambda \vert m_z \vert\lambda \rangle = H_S+\Gamma (2
n_\lambda/N-1) |\uparrow\rangle\langle\uparrow | $, where $n_\lambda $
is the fermionic occupation number of $|\lambda\rangle $.

Let us start analyzing the zero-temperature scenario, where the bath
is in its ground state $|G\rangle$, which is characterized by its
fermionic occupation number $n_G$, which in turn depends on $h$. In
this case, the sum over all the Hamiltonian eigenstates in
Eq.~(\ref{wei}) reduces to only one term, $\rho_{S} ^{ij}(t)= \langle
i\vert e^{-i H_S^{(G)}t}\rho_S (0) e^{i H_S^{(G)} t} \vert j \rangle$
and the transition probability between $| \downarrow \rangle$ and $|
\uparrow \rangle$ is given by $P_{tr}=\left|\langle \downarrow |e^{-i
    H_S^{(G)}t}\rho_S (0) e^{i H_S^{(G)} t} |\uparrow
  \rangle\right|^2$.  This allows to renormalise the diagonal elements
of the system Hamiltonian and the effective gap between $| \uparrow
\rangle$ and $| \downarrow \rangle$ becomes $\tilde{\Delta}=
\Delta-\Gamma(1-2 n_{G}/N)$.  For $h=0$, 
where $n_G=N/2$, the presence of the bath does not affect $P_{tr}$,
but for increasing field $h$ the fermionic occupation number starts
decreasing until we have $n_G=0$ at $h=h_c=1$. The value of $n_G/N$ in
the thermodynamic limit for $\gamma \to 0 $ and $\beta \to \infty$ can
be obtained from Eq.~(\ref{magn}).
If $\Gamma \ge \Delta$ it is therefore possible to achieve perfect
energy transfer by choosing an optimal $h$ between zero and one such
that $\tilde{\Delta}=0$. For the case of $\Gamma < \Delta$, which is
the experimentally more realistic scenario, a maximum enhancement will
be reached for $h\ge h_c$, which corresponds to the zero-temperature
case described in Ref.~\cite{sinayskiy}.

\begin{figure}[t]
\begin{center}
\includegraphics[width=8cm]{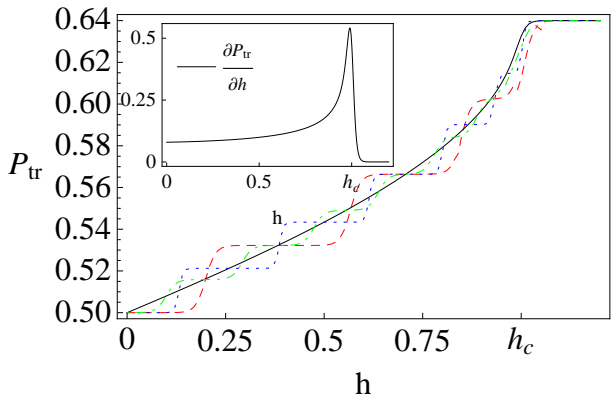}
  \caption{(Color online) $P_{tr}$ versus $h$ for $N=8$ sites [red
    (dashed) line], $N=12$ sites [blue (dotted) line], $N=16$ sites
    [green (dot-dashed) line], and for $N=\infty$ sites [black (solid)
    line], in the isotropic case ($\gamma=0$). The system parameters
    are $\Delta=2T=0.2$ and $\Gamma=5\times 10^{-2}$, while the
    inverse bath temperature is $\beta=40$. In the case of $8$, $12$,
    and $16$ sites, $P_{tr}$ has been calculated by using the
    analytical expression given in Eq.~(\ref{wei}), while the
    thermodynamic limit has been calculated within a mean-field
    approximation (see text for details). In the absence of the bath,
    one would have $P_{tr}=0.5$. Inset: the derivative of $P_{tr}$
    with respect to $h$ in the thermodynamic limit. For finite-size
    chains, the derivatives would exhibit $N/2$ peaks, the last of
    them falling around $h_c$, in correspondence with the degeneracy
    points of the Hamiltonian.}
\label{fig1} 
\end{center}
\end{figure}

For systems at finite temperatures we need to consider all terms of
the sum in Eq.~(\ref{wei}), which, for any finite $N$, can be
evaluated analytically. For a finite inverse temperature $\beta=40$,
we show $P_{tr}$ in Fig.~\ref{fig1} as a function of $h$ for chains of
8, 12 and 16 spins and compare it to its value in the thermodynamic
limit. The later is calculated by introducing a mean-field
approximation, obtained by replacing $H_{I}$ with
$H_{I}^\text{mf}=-\Gamma |\uparrow\rangle\langle\uparrow | \langle m_z
\rangle$, where the average magnetization per spin is given by
Eq.~(\ref{magn}) in the limit of vanishing $\gamma$.  By comparing the
different curves, it is evident that the curves for finite-sized
chains are rapidly converging toward the mean-field approximation,
which can then be used to estimate $P_{tr}$ in the limit of long
chains. The finite-size solutions show clear signatures of the $N/2$
transition points in the ground state between the different symmetry
sectors of the Hamiltonian and exhibit horizontal plateaus between
them. Taking the thermodynamic limit, where these transition points
become infinitely dense, we are left with a critical change for
$P_{tr}$ at $h\approx h_c$. The existence of a critical point in
$P_{tr}$ is shown in the inset of Fig.~\ref{fig1}, where its
derivative with respect to the external field is plotted. It shows a
clear peak around $h_c$, which is due to the behavior of $m_z$.

In the zero-temperature case, the behavior of $P_{tr}$ around $h_c$ is
a direct consequence of the commutativity between the bath and the
interaction Hamiltonians. 
Indeed, the only effect of the bath is to replace the initial,
unperturbed, system Hamiltonian $H_S$ with the effective two-level
Hamiltonian $H_S^{(G)}$, which has acquired the critical properties of
the transverse magnetization.
However, in the finite-temperature case, this approach does not hold
any more. Nevertheless, the results of Fig.~\ref{fig1} show that, for
moderate temperatures, which can be expected in realistic scenarios, a
connection between  $P_{tr}$ and the critical change in $m_z$ still
holds (further decreasing $\beta$ would lead the bath out of its
quantum domain and wash out the effect on $P_{tr}$). We will discuss
this point in more detail below.

\begin{figure}
\begin{center}
\includegraphics[width=8cm]{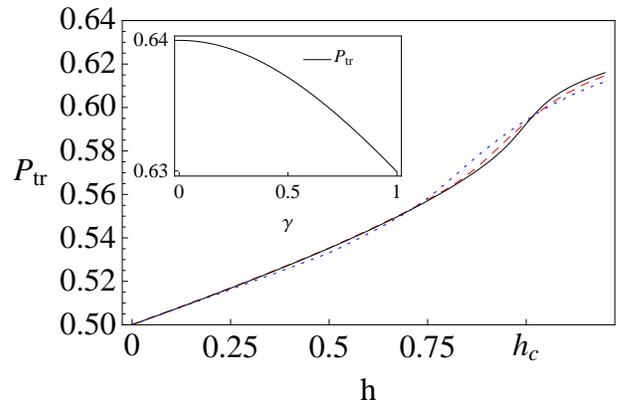}
  \caption{(Color online) $P_{tr}$ versus $h$ for $N=6$ sites [blue
    (dotted) line], $N=10$ sites [red (dashed) line], and for
    $N=\infty$ sites [black (solid) line], in the presence of an
    anisotropic bath ($\gamma=0.8$). The system parameters are
    $\Delta=2T=0.2$ and $\Gamma=5*10^{-2}$, while the inverse bath
    temperature is $\beta=40$. Here, the finite-size lines have been
    calculated by explicitly solving the Hamiltonian evolution, while
    the mean-field approximation has been used in the thermodynamic
    limit. Inset: $P_{tr}$ as a function of $\gamma$ for strong fields
    ($h=2$) and for $\beta=40$. The transition probability is a
    monotonically decreasing function of the anisotropy.
    }
\label{fig2} 
\end{center}
\end{figure}

Releasing the isotropy assumption by choosing $\gamma \neq 0$ in
Eq.~(\ref{hxy}), an analytical solution for $\rho_{S} ^{ij}(t)$ analogous to that
in Eq.~(\ref{wei}) cannot be obtained. Instead, the general expression
$P_{tr} =\max_t \left| \langle \downarrow |{\rm Tr}_B [ e^{-i
    H_\text{tot} t}\rho_S (0)\otimes \rho_B(0) e^{i H_\text{tot} t}
  ]|\uparrow \rangle \right|^2$,
where $H_\text{tot}=H_{S}+H_{XY}+H_{I}$, must be evaluated. In the
case of short chains, this can be done exactly by diagonalizing
$H_\text{tot}$ and explicitly performing the trace over the
environment. In Fig.~\ref{fig2} we show $P_{tr}$ for chains of 6 and
10 spins at finite temperature and compare the results to the
mean-field approximation, performed, as before, by replacing $m_z$
with its average value in $H_{I}$. 
One can see that the mean field solution is a good approximation to
the exact, finite-size solution even for short chains, and therefore
becomes a suitable candidate to describe the thermodynamic
limit. Furthermore, this behavior shows that the influence of a
quantum phase transition in the bath on the transition dynamics
between the two internal system states is not limited to a special
choice of Hamiltonian or to possible special commutation properties,
but lies in nature of the interaction.

\begin{figure}
\begin{center}
\includegraphics[width=8cm]{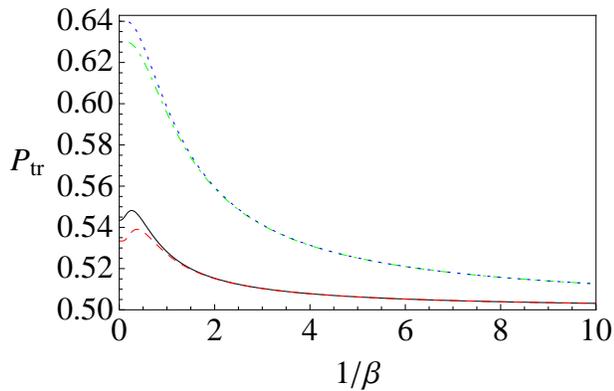}
  \caption{(Color online) $P_{tr}$ versus $1/\beta$, in the
    thermodynamic limit, for $\gamma=0$ and $h=2$ [blue (dotted)
    line], $\gamma=1$ and $h=2$ [green (dotdashed) line], $\gamma=0$
    and $h=0.5$ [black (solid) line], and $\gamma=1$ and $h=0.5$ [red
    (dashed) line].}
\label{fig3} 
\end{center}
\end{figure}

Let us finally discuss the interplay between temperature effects and
anisotropy for both small ($h=0.5$) and large fields ($h=2$). In
Fig. \ref{fig3} we show $P_{tr}$ as a function of $\beta$ for the
isotropic case $\gamma=0$ and for the Ising case $\gamma=1$. Two
remarkable results emerge: i) the bath can have a positive effect on
$P_{tr}$ for a large range of temperatures and ii) $\gamma$ is in fact
important only in the very low temperature regime. The latter result
not only holds for strong fields, where the $x-y$ component of the
interaction is dominated by the external field, but also in the
symmetry-broken region, where the in-plane and the transverse terms
have comparable strength.

In conclusion, we have considered the problem of a  TLS
immersed in a spin chain that plays the role of the environment. We
have shown that the transition probability  from the
lower level to the upper level of the  TLS  is sensitive to the
critical properties of the bath. Describing the bath as an isotropic
XX chain and the interaction with the system through its transverse
magnetization, the problem becomes exactly solvable, since the bath
Hamiltonian commutes with the interaction term. In the
zero-temperature case, the bath renormalizes the system Hamiltonian
making it possible to improve the transition probability. Even if the
mechanism is more complicated, the same qualitative enhancement can be
observed in the finite temperature regime. Finally, the
decoherence-assisted process identified in our work has a more
general range of validity, since we have shown that the same
qualitative effect can be observed by using an anisotropic chain,
where the commutativity property between bath and interaction
Hamiltonian does not apply any more.

 We believe that our results can elucidate the interplay between quantum coherence and correlations and can help understand the basic mechanisms underlying  energy transport in biologic systems.

This work was supported by Science Foundation of Ireland under project
number 10/IN.1/I2979. We would like to thank Tony Apollaro, Mauro
Paternostro, Gabriele De Chiara, and Ferdinando de Pasquale for
valuable discussions and comments.


\begin{thebibliography}{0}

\bibitem{nielsen}  M. A. Nielsen and I. L. Chuang,  {\it Quantum Computation and Quantum Information},  (Cambridge University Press, Cambridge, 2000).

\bibitem{breuer} H.-P. Breuer and F. Petruccione, {\it The Theory of Open Quantum Systems} (Oxford University Press, New York, 2002).

\bibitem{engel2007} G. S. Engel, T. R. Calhoun, E. L. Read, T.-K. Ahn, T.~Man{\v c}al, Y.-C. Cheng, R. E. Blankenship, and G. R. Fleming,
 Nature (London)  {\bf 446}, 782 (2007).

\bibitem{panit2010} G. Panitchayangkoon, D. Hayes, K. A. Fransted, J. R. Caram, E. Harel, J. Wen, R. E.
  Blankenship, and G. S. Engel, Proc. Natl. Acad. Sci. {\bf 107}, 12766 (2010).

\bibitem{collini2010} E. Collini, C. Y. Wong, K. E. Wilk, P. M. G. Curmi, P. Brumer,
  and G. D. Scholes, Nature (London)  {\bf 463},  644 (2010).


\bibitem{plenio08}
M. Plenio and S. Huelga,  New J. Phys. {\bf 10}, 113019, (2008).

\bibitem{olaya08}
A. Olaya-Castro, C. Lee, F. Olsen, and N. Johnson, Phys. Rev. B  {\bf 78}, 085115 (2008).


\bibitem{mohseni08}
M. Mohseni, P. Rebentrost, S. Lloyd, and A. Aspuru-Guzik, J. Chem. Phys. {\bf 129}, 174106 (2008).



\bibitem{rebentrost09}
P. Rebentrost, M. Mohseni, I. Kassal, S. Lloyd, and  A.  Aspuru-Guzik, New J. Phys. {\bf 11}, 033003 (2009).



\bibitem{caruso09}
F. Caruso, A. W. Chin, A. Datta, S. F. Huelga, and M. B. Plenio, J. Chem. Phys. {\bf 131}, 105106 (2009); Phys. Rev. A  {\bf 81}, 062346 (2010).



\bibitem{sarovar} M. Sarovar,	 A. Ishizaki,	 G. R. Fleming, and	  K. B. Whaley, Nature Phys. \textbf{6}, 462 (2010).

\bibitem{semiao} F. L.  Semi{\~a}o, K. Furuya, and G. J. Milburn, New J. Phys. {\bf 12}, 083033 (2010).



\bibitem{campos11}
L. Campos Venuti and P. Zanardi,  Phys. Rev. B  {\bf 84}, 134206 (2011).


\bibitem{kassal12}
I. Kassal and A.  Aspuru-Guzik, New J. Phys. {\bf 14}, 053041 (2012).


\bibitem{sinayskiy} I. Sinayskiy, A. Marais, F. Petruccione, and A. Ekert,  Phys. Rev. Lett. {\bf 108}, 020602 (2012).


\bibitem{paganelli}  S. Paganelli, F. de Pasquale, and S. M. Giampaolo,  Phys. Rev. A  {\bf 66}, 052317 (2002).

\bibitem{quan}  H. T. Quan, Z. Song, X. F. Liu, P. Zanardi, and C. P. Sun, Phys. Rev. Lett. {\bf 96}, 140604 (2006).


\bibitem{rossini07}  D. Rossini, T. Calarco, V. Giovannetti, S. Montangero, and R. Fazio,
  Phys. Rev. A {\bf 75}, 032333 (2007); J. Phys. A: Math. Theor. {\bf 40}, 8033 (2007).

  


\bibitem{rossini2008}  D. Rossini, P. Facchi, R. Fazio, G. Florio, D. A. Lidar, S. Pascazio,  F. Plastina, and P. Zanardi,  
  Phys. Rev. A {\bf 77}, 052112 (2008).
  
\bibitem{ions} G. L. Giorgi, S. Paganelli, and F. Galve, Phys. Rev. A {\bf 81}, 052118 (2010).
  
\bibitem{damski} B. Damski, H. T. Quan, W. H. Zurek, Phys. Rev. A {\bf 83}, 062104 (2011).


\bibitem{dawson} C. M. Dawson, A. P. Hines, R. H. McKenzie, and G. J. Milburn,   Phys. Rev. A {\bf 71}, 052321 (2005).
   
\bibitem{barouch} E. Barouch, B. M. McCoy, and M. Dresden, Phys. Rev. A {\bf 2}, 1075 (1970).
  
\bibitem{lieb}  E. Lieb, T. Schultz and D. Mattis, Ann. Phys. (N.Y.) {\bf 16}, 407 (1961).


\end{thebibliography}
\end{document}